\documentclass[varenna]{cimento}

\usepackage{epsfig}

\title{Mesoscopic quantum measurements}

\author{Dmitri V. Averin}

\institute{Department of Physics and Astronomy, \\ Stony Brook
University, SUNY \\ Stony Brook, NY 11794-3800, USA}

\shortauthor{D.V. Averin}

\begin{document}

\maketitle

\begin{abstract}
The paper discusses dynamics of quantum measurements in mesoscopic
solid-state systems. The aim is to show how the general ideas of
the quantum measurement theory play out in the realistic models of
actual mesoscopic detectors. The two general models of ballistic
and tunneling detectors are described and studied quantitatively.
Simple transformation cycle demonstrating wavefunction reduction
in a mesosocpic qubit is suggested.

\end{abstract}

\section{Introduction}

Despite the long history of quantum mechanics, the quantum
measurement problem continues to attract interest motivated mostly
by the counter-intuitive features of the ``wavefunction
reduction''. Although dynamics of any measurement set-up is
governed by the Schr\"{o}dinger equation with an appropriate
Hamiltonian, full description of the measurement process can not
be obtained without account of the changes in the wavefunction of
the measured system caused by the random process of selection of
one specific outcome of measurement out of the range of possible
outcomes. This selection process is trivial in the case of
classical dynamics, when all possible outcomes of measurement are
``orthogonal'' and the observation of the measured system in one
particular state does not imply any changes in the system beyond
the statement that it occupies this and not any other state. For a
quantum system, however, existence of the non-commuting
observables implies that selection of one particular outcome of
measurement can change the wavefunction of the system in a highly
non-trivial way. Such a reduction of the wavefunction appears as
an evolution principle additional to the Schr\"{o}dinger equation.
Moreover, the changes in the wavefunction of the measured system
induced by it can violate the basic features of dynamics which
follows from the Schr\"{o}dinger equation, despite the fact that
the measurement process as a whole is governed by this equation.
The best known example of this situation is the case of EPR
correlations \cite{r1} between the two spatially separated spins,
which violate the no-action-at-a-distance principle as quantified
by the Bell's inequalities \cite{r2}. From the perspective of the
wavefunction reduction, the EPR correlations appear as a result of
selection of one random specific outcome of the local spin
measurement. On average, there is no action-at-a-distance in a
sense that the correlations by themselves can not lead to
information transfer between the points where the spins are
located.

Current interest to the solid-state quantum information processing
(see, e.g., the reviews \cite{l3,l4,l5}), motivates development of
mesoscopic solid-state structures that can serve both as simple
quantum systems, e.g., qubits or harmonic oscillators, and the
detectors. Although in experiments, the mesoscopic detectors did
not reach the stage yet where they can be used to look into the
basic questions of the quantum measurement theory (which requires
the quantum-limited detection) one can expect this to happen quite
soon. A new element introduced by the mesoscopic structures in the
discussion of quantum measurements is the fact that the
wavefunction reduction is not necessarily caused by interaction of
a ``microscopic'' measured system with the ``macroscopic''
detector. In mesoscopic structures, the measured systems and the
detectors are similar in many respects (including dimensions and
typical dynamics) and are frequently interchangeable: a measured
system in one context can act as the detector in the other, and
vice versa. This shows that the boundary at which the quantum
coherent dynamics should be complemented with the wavefunction
reduction is not universal.

The aim of this work is to provide a quantitative discussion of
models and measurement dynamics of the mesoscopic detectors. The
discussion emphasizes the interplay between the dynamic and
information sides of the measurement process and can serve as an
introduction to the problem of wavefunction reduction in
mesoscopic structures.

\section{Measurements dynamics of ballistic mesoscopic detectors}

\begin{figure}
\hspace{4cm} \epsfxsize=5cm \epsfbox{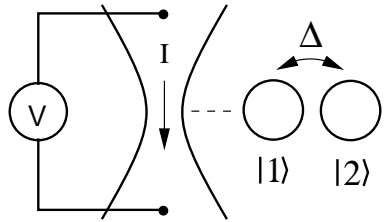} \caption{Schematic
of the QPC detector measuring charge qubit. The two qubit states
$|j\rangle$, $j=1,2$ are localized on the opposite sides of a
tunnel barrier and are coherently coupled by tunneling across this
barrier with coupling strength $\Delta$. Transfer of the qubit
charge between the states $|j\rangle$ changes electrostatic
potential in the scattering region of the QPC affecting the
current $I$ through it that is driven by the applied voltage $V$.}
\label{fm2}
\end{figure}

Majority of the mesoscopic detectors use as their operating
principle ability of a measured system to control the transport of
some particles between the two reservoirs. The information about
the state of the system is contained then in the magnitude of the
particle current between the reservoirs which serves as the
detector output. In the most direct form, this principle is
implemented in the quantum point contact (QPC)
detector~\cite{q1,q2}, which presently is the main detector used
for measurements of the quantum dot qubits~\cite{q3,q4,q5,q6,q7}.
In the QPC detector (Fig.~\ref{fm2}), the propagating particles
are electrons which move ballistically through a short
one-dimensional constriction formed between the two electrodes of
the QPC. The electrodes can be viewed as reservoirs of independent
and effectively non-interacting electrons. The measured system
creates electrostatic potential that makes the scattering
potential $U_j(x)$ for electrons in the constriction dependent on
the state $|j\rangle$ of the system, and in this way controls
transmission probability of the QPC. The output of the QPC
detector is the electric current $I$ driven by the voltage
difference $V$ between the electrodes. The current depends on the
electron transmission probability, and as a result, contains
information about the state $|j\rangle$. Since interaction between
the QPC electrons and the measured system is dominated by the
electrostatic potential, the QPC acts as the charge detector.
Another example of the ballistic mesoscopic detector is the
magnetic analog of the QPC based on the ballistic motion of the
magnetic flux quanta (fluxons) through a one-dimensional channel,
the role of which is played by the Josephson transmission line
(JTL)~\cite{q8}. The scattering potential $U_j(x)$ for the fluxons
in the JTL is created by the magnetic flux or current, and the JTL
detector can be used for measurements of superconducting flux
qubits. In the JTL detector, the fluxons can be injected into the
JTL individually providing control over the individual scattering
events.

The detector model in which the output information is contained in
the transport current flowing between the two reservoirs applies
to many of the mesoscopic detectors (see Sec. 4). There are
several reasons for this. One is the strong (in the tunnel limit,
exponential) dependence of the scattering amplitudes on parameters
of the scattering potential that leads to sufficiently large
sensitivity of the detector to the measured system. Another, more
important, is the fact that the scattering dynamics contains
strongly divergent transmitted and reflected trajectories that
create easily detectable different outcomes of measurement. This
feature of scattering is not easily reproducible in other types of
the dynamics~\cite{q9}. Finally, the transport between large
reservoirs makes it possible to repeat scattering events at a
certain rate amplifying the results of scattering of one particle.

\begin{figure}
\hspace{2.5cm} \epsfxsize=8cm \epsfbox{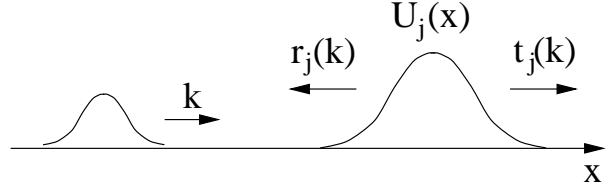}
\caption{Measurement dynamics of a ballistic mesoscopic detector.
The wavepacket of a particle with momentum $k$ is scattered by the
potential $U_j(x)$ controlled by the measured system. The
scattering potential and the transmission/reflection amplitudes
$t_j(k)$, $r_j(k)$ contain information about the state $|j\rangle$
of the system.} \label{fm1}
\end{figure}

In general, the process of quantum measurement can be understood
as creation of an entangled state of the measured system and the
detector as a result of interaction between them. The states of
the detector are classical and suppress quantum superposition of
different outcomes of measurement. The two consequences of this
process are the acquisition of information about the system by the
detector, and ``back-action'' dephasing of the measured system -
see, e.g., \cite{q10,q11}. Because of the system-detector
entanglement, finding a given detector output provides some
indication of what state the measured system is in. On the other
hand, the same entanglement means that quantum coherence among the
states of the measured system is suppressed. This implies that
there is a close connection between the information acquisition
and back-action dephasing. In the optimal situation, the rate $W$
with which the detector obtains information about the system and
the dephasing rate $\Gamma$ are the same. Of course, the detector
can always introduce some parasitic dephasing into the system
dynamics, so that in general $W\leq \Gamma$. In view of this
inequality, the detector with $W= \Gamma$ is called ``ideal'' or
``quantum limited''. If the detector is far from being
quantum-limited, it destroys quantum coherence in the measured
system long before it provides information that can be used to
select specific outcomes of measurement. Because of this, only the
detectors that are close to being quantum-limited can give rise to
non-trivial wavefunction reduction.

\subsection{Back-action dephasing rate}

Measurement dynamics with the ballistic mesoscopic detector is
illustrated in Fig.~\ref{fm1}. For the ballistic detector, the
detector-system entanglement arises as a result of scattering
(Fig.~\ref{fm1}), and the rates of information acquisition and
back-action dephasing can be expressed in terms of the scattering
amplitudes \cite{q12}. To do this, we consider evolution of the
density matrix $\rho$ of the measured system in scattering of one
particle. For simplicity, the Hamiltonian of the system itself is
assumed to be zero (e.g., $\Delta=0$ in the example of
Fig.~\ref{fm2}), and the system evolution is caused only by the
interaction with the detector. The evolution of $\rho$ is obtained
then from the time dependence of the total wavefunction of the
scattered particle and the stationary wavefunction $\sum_j c_j
|j\rangle$ of the measured system:
\begin{equation}
\psi(x,t=0) \cdot \sum_j c_j|j\rangle \rightarrow \sum_j
c_j\,\psi_j(x,t) \cdot |j\rangle \, . \label{e11}
\end{equation}
Here $\psi(x,t=0)$ is the initial wavefunction of the particle
injected in the scattering region from the reservoir, and its time
evolution $\psi_j(x,t)$ depends on the realization $U_j(x)$ of the
potential created by the measured system. Tracing over the
detector, i.e., the scattering wavefunction, one gets from
Eq.~(\ref{e11}):
\begin{equation}
\rho_{ij} =c_ic_j^* \rightarrow c_ic_j^* \int dx \psi_i(x,t)
\psi_j^*(x,t) \, . \label{e12}
\end{equation}

Qualitatively, the time evolution in (\ref{e11}) describes
propagation of the initial wavepacket towards the scattering
potential and its subsequent separation in coordinate space into
the transmitted and reflected parts that are well-localized on the
opposite sides of the scattering region. At time $t>t_{sc}$, where
$t_{sc}$ is the characteristic scattering time, the separated
wavepackets move in the region free from the $j$-dependent
potential and the unitarity of the quantum-mechanical evolution of
$\psi_j(x,t)$ implies that the overlap of the scattered
wavefunctions in Eq.~(\ref{e12}) becomes independent of $t$. This
overlap can be directly found in the momentum representation:
\begin{equation}
\int dx \psi_i(x,t) \psi_j^*(x,t)= \int dk
|b(k)|^2[t_i(k)t_j^*(k)+r_i(k) r_j^*(k)]\, , \label{e13}
\end{equation}
where $b(k)$ is the probability amplitude for the injected
particle to have momentum $k$ in the initial state. Equations
(\ref{e12}) and (\ref{e13}) show that the diagonal elements of the
density matrix $\rho$ do not change in the scattering process:
\begin{equation}
\int dk |b(k)|^2[|t_i(k)|^2+|r_i(k)|^2]=1 \, , \label{e14}
\end{equation}
while the off-diagonal elements are suppressed by the factor
\begin{equation}
\left|\int dk |b(k)|^2 [t_i(k)t_j^*(k)+ r_i(k)r_j^*(k)]
\right|\leq 1 \, .\label{e14*}
\end{equation}

The inequality in this relation follows from the Swartz inequality
for the scalar product in the Hilbert space of ``vectors''
$|b(k)|\cdot \{t_j(k),r_j(k)\}$ of the unit length (\ref{e14}).
Suppression of the off-diagonal elements of $\rho$ is
manifestation of the back-action dephasing of the measured system
by the detector. Assuming that the particles are injected from the
reservoir with frequency $f$ and combining the suppression factors
(\ref{e14*}) for the successive scattering events, we obtain the
dephasing rate as
\[ \Gamma_{ij} = - f \ln \left|\int dk |b(k)|^2 [t_i(k)t_j^*(k)+
r_i(k)r_j^*(k)] \right| \, . \] If the scattering amplitudes do
not depend on momentum $k$ in the range of momenta limited by
$|b(k)|^2$, the back-action dephasing rate becomes independent of
the form of initial wavepacket of injected particles:
\begin{equation}
\Gamma_{ij} = - f \ln \left| t_it_j^*+ r_ir_j^* \right| \, .
\label{e15}
\end{equation}
Equation (\ref{e15}) is the general expression for the back-action
dephasing rate of a ballistic mesoscopic detector with momentum-
(and energy-) independent scattering amplitudes. It is valid, in
particular, for the QPC detector at low temperatures $T\ll eV$, if
the injection frequency $f$ is taken to be equal to the ``attempt
frequency'' $eV/h$ with which electrons are incident on the
scattering region~\cite{q13}.

In the linear-response regime, when the changes in the scattering
amplitudes with $j$ are small, the limiting form of
Eq.~(\ref{e15}) was obtained in \cite{l6,l7,b9,l8,l9}. In the
tunnel limit $t_j\rightarrow 0$, Eq.~(\ref{e15}) reduces to
\begin{equation}
\Gamma_{ij} = (f/2)| \bar{t}_i-\bar{t}_j |^2 \, , \label{e151}
\end{equation}
where
\begin{equation}
\bar{t}_j \equiv |t_j|e^{i\phi_j} , \;\; \phi_j \equiv
\mbox{arg}(t_j /r_j) \, . \label{e152}
\end{equation}
When the phases $\phi_j$ can be neglected, Eq.~(\ref{e151})
reproduces earlier results \cite{gur} for the dephasing by the QPC
detector in the tunnel limit. As we will see in Sec.~4,
Eq.~(\ref{e151}) can be obtained in this limit under more general
assumptions, and describes large number of different mesoscopic
detectors.

\subsection{Information acquisition rate}

As was discussed above, the back-action dephasing is only one part
of the measurement process. The other part is acquisition by the
detector of information about the state of the measured system. In
the case of ballistic detector, the information is contained in
the scattering amplitudes of the incident particles, and the rate
of its acquisition depends on specific characteristics of the
amplitudes recorded by the detector. One of the simplest
possibilities in this respect, realized, e.g., in the QPC
detectors, is to record the changes in the transmission
probabilities which determine the magnitude of the particle
current through the scattering region. (Alternatively, one could
modify the scattering scheme by forcing the scattered particles to
interfere, and in this way use the phase information \cite{q12} in
the scattering amplitudes.) The rate of information extraction
from the current magnitude, i.e., the rate of increase of the
confidence level in distinguishing different states $|j\rangle$,
can be calculated simply by starting with the
transmission/reflection probabilities $T_j=|t_j|^2$ and
$R_j=1-T_j$ when the measured system is in the state $|j\rangle$.
Since successive scattering event are independent, the probability
$p_j(n)$ to have $n$ out of $N$ incident particles transmitted, is
given by the binomial distribution $p_j(n)=C_N^n T_j^nR_j^{N-n}$.
The task of distinguishing different states $|j\rangle$ of the
measured system is transformed by the detector into distinguishing
the probability distributions $p_j(n)$ for different $j$s. Since
the number $N=ft$ of scattering attempts increases with time $t$,
the distributions $p_j(n)$ become peaked successively more
strongly around the corresponding average numbers $T_jN$ of
transmitted particles. The states with different probabilities
$T_j$ can be distinguished then with increasing certainty. The
rate of increase of this certainty can be characterized
quantitatively by some measure of the overlap of the distributions
$p_j(n)$. While in general there are different ways to
characterize the overlap of different probability distributions
\cite{q14}, the characteristic which is appropriate in the quantum
measurement context \cite{q15,q11} is closely related to
``fidelity'' in quantum information \cite{q14}: $\sum_n
[p_i(n)p_j(n)]^{1/2}$. The rate of information acquisition can
then be defined naturally as \cite{q12}:
\begin{equation}
W_{ij} = - (1/t) \ln \sum_n [p_i(n) p_j(n)]^{1/2} \, . \label{e17}
\end{equation}
Using the binomial distribution in this expression we get:
\begin{equation}
W_{ij} = - f \ln [(T_iT_j)^{1/2}+(R_iR_j)^{1/2} ] \, . \label{e18}
\end{equation}

Equation (\ref{e18}) gives the information acquisition rate by
ballistic mesoscopic detectors. Comparing Eqs.~(\ref{e15}) and
(\ref{e18}) we see that for this type of the detector, in
accordance with the general understanding of quantum measurements,
the back-action dephasing rate and information acquisition rate
satisfy the inequality
\begin{equation}
W_{ij} \leq \Gamma_{ij} \, . \label{e21}
\end{equation}
Equality in this relation gives the condition of the
quantum-limited operation of the ballistic mesoscopic detector
under the assumption of energy-independent scattering amplitudes.
It holds if
\begin{equation}
\phi_j = \phi_i \, , \label{e19}
\end{equation}
where the phases $\phi_j$ are defined in Eq.~(\ref{e152}).
Condition (\ref{e19}) has simple interpretation as the statement
that there is no information on the states $|j\rangle$ in the
phases of the scattering amplitudes. Deviations from
Eq.~(\ref{e19}) mean that the phases contain information about the
measured system which is lost in the detection scheme sensitive
only to the transmission probabilities $T_j$. In this case, the
information loss in the detector prevents it from being
quantum-limited. In practical terms, the simplest way to satisfy
condition (\ref{e19}) is to make the scattering potential
symmetric $U_j(-x)=U_j(x)$ for all states $|j\rangle$. The
unitarity of the scattering matrix for this potential implies then
that $\phi_j =\pi/2$ for any $j$, and Eq.~(\ref{e19}) is
automatically satisfied. If the detector is quantum-limited, it
can demonstrate non-trivial wavefunction reduction.

\subsection{Conditional evolution}

Quantitative description of the wavefunction reduction due to
interaction with a detector can be formulated as ``conditional''
evolution, in which dynamics of the measured system is conditioned
on the observation of particular outcome of measurement. In the
axiomatic approach, wavefunction reduction is formalized together
with the dynamic evolution as ``quantum operation''~\cite{r3},
arbitrary linear transformation of the system density matrix
satisfying physically motivated axioms. In this approach, a
detector is characterized by a set of positive operators which
correspond to all possible outcomes of measurements with this
detector, or ``positive operator valued measure''
(POVM)~\cite{r4}. In practice, for any real specific detector, it
is clear what the possible classical outcomes of measurements are,
and the emphasis is then on development of dynamic equations that
would describe evolution of the measured system conditioned on a
given detector output. Since the different outcomes of the
detector evolution are classically distinguishable, it is
meaningful to ask how the measured system evolves for a given
output. Such a conditional evolution of the measured system
describes quantitatively the wavefunction reduction in the
measurement process with a particular detector (see, e.g.,
\cite{q16,q17,q18,q11}). In the case of ballistic mesoscopic
detectors, each act of particle scattering represents an
elementary measurement process. Since the particle trajectories
that correspond to different outcomes of scattering: transmission
through or reflection from the scattering region are strongly
separated, these outcomes should be considered as non-interfering
classical events. Although the absence of quantum coherence
between these two outcomes is an assumption of the conditional
approach, this assumption is very natural. Propagation of the
scattered particles in different reservoirs of the detector
entangles them with different environments, the process that very
efficiently suppresses their mutual quantum coherence \cite{r5}.
While this ``common-sense'' assumption of absence of quantum
coherence between different outputs of a realistic detector is
sometimes considered unsatisfactory from an abstract point of view
\cite{r6,l1}, it can be given a fairly rigorous description in
terms of decoherence in open quantum systems -- see, e.g.,
\cite{r7}.

Quantitatively, conditional equations are obtained by separating
in the total wavefunction the terms that correspond to a specific
classical outcome of measurement and renormalizing this part of
the wavefunction so that it corresponds to the total probability
of 1~\cite{q20,q11,q8}. In the ballistic detector, there are two
classically different outcomes of scattering, transmission and
reflection, for each injected particle. This means that the
wavefunction of the measured system should be conditioned on the
observation of either transmitted or reflected particle in each
elementary cycle of measurement. The evolution of the total
wavefunction ``detector+measured system'' as a result of
scattering of one particle is described by Eq.~(\ref{e11}). Under
the assumption of energy-independent scattering amplitudes,
momentum and  coordinate dependence of the states of the scattered
particles in the detector is the same for different states
$|j\rangle$, and can be factored out from the total wavefunction.
The evolution of the measured system can then be conditioned on
the transmission/reflection of a particle simply by keeping in
Eq.~(\ref{e11}) the terms that correspond to the actual outcome of
scattering in the form of the appropriate scattering amplitudes.
If the particle is transmitted through the scattered region or
reflected from it in a given measurement cycle, amplitudes $c_j$
for the system to be in the state $|j\rangle$ change then,
respectively, as follows:
\begin{equation}
c_j \rightarrow t_j c_j \big/ \left[\Sigma_j |c_j |^2
T_j\right]^{1/2}, \;\;\;\; c_j \rightarrow r_j c_j \big/
\left[\Sigma_j |c_j |^2 R_j \right]^{1/2} . \label{e32}
\end{equation}
We see that the expansion coefficients $c_j$ of the system's
wavefunction are changing in conditional evolution despite the
initial assumption that the system Hamiltonian is zero. This is
unusual from the point of view of the Schr\"{o}dinger equation,
and provides quantitative expression of reduction of the
wavefunction in the measurement process.

Also, it should be noted that the transformations (\ref{e32}) do
not decohere the measured system despite the back-action dephasing
by the detector discussed above. To understand this, one should
note that as in the case of any dephasing, the back-action
dephasing can be viewed as the loss of information. For the
quantum-limited detection, the overall evolution of the detector
and the measured system is quantum-coherent and the only possible
source of the information loss is averaging over the detector
evolution. This means that the back-action dephasing arises as the
result of averaging over different measurement
outcomes~\cite{q19}, and specifying definite outcome as done in
the conditional dynamics removes all losses of information and
eliminates the dephasing.

Equations (\ref{e32}) can be applied directly to the detectors
which provide control over scattering of individual particles,
e.g. to the JTL detector~\cite{q8}, where it makes sense to
discuss changes in the wavefunction of the measured system induced
by one scattering event. In some detectors, however, such a
control over individual scattering events is not fully possible.
For instance, in the QPC detector, the picture of individual
scattering events leading to Eqs.~(\ref{e15}) and (\ref{e18}) for
the back-action dephasing and information rates is strictly
speaking valid only on the relatively large time scale $t \gg
h/eV$, when the typical number of electron scattering attempts is
larger than 1. One can generalize Eq.~(\ref{e32}) to this
situation by considering the time interval $t$ which includes a
number $N=ft>1$ of scattering attempts, where for the QPC
$f=eV/h$. Combining the transformations (\ref{e32}) one can see
that observation of any sequence of transmission/reflection events
that includes $n$ transmissions and $N-n$ reflections changes the
wavefunciton as:
\begin{equation}
c_j \rightarrow t_j^n r_j^{(N-n)} c_j \Big/ \left[\Sigma_j |c_j|^2
T_j^n R_j^{(N-n)}\right]^{1/2},   \label{e33}
\end{equation}
regardless of the specific ordering of these events. This equation
includes as a particular case Eq.~(\ref{e32}) which follows when
$N=1$. Since the wavefunction obtained as a result of
transformation (\ref{e33}) is the same for all $C_N^n$ sequences
with the same total number $n$ of transmitted particles, one can
distinguish all scattering outcomes only by $n$. For each of the
$N+1$ outcomes with different $n$ the wavefunction is transformed
according to Eq.~(\ref{e33}). This means that the wavefunction
reduction has the form (\ref{e33}) independently of whether the
detector suppresses quantum coherence between all the sequences of
scattering outcomes or only between the states with different
total numbers of transmitions/reflections. Transformations
(\ref{e33}) can be used to study quantitatively unusual
manifestations of the wavefuntion reduction in the mesoscopic
solid-state qubits.

\section{Tunneling without tunneling: wavefunction reduction in a
mesoscopic qubit}

Probably the simplest example of the counter-intuitive features of
the wavefuntion reduction in mesoscopic qubits arises from a
question whether a quantum particle can tunnel through a barrier
which has vanishing transparency? Immediate answer to this
question is ``no'' as follows from the elementary properties of
the Schr\"{o}dinger equation. It seems basically the tautology to
say that if the tunneling amplitude is zero (e.g., the barrier is
infinitely high) the tunneling is suppressed. Of course somewhat
more careful consideration reminds that evolution according to the
Schr\"{o}dinger equation is not the only way for a state of a
quantum particle to change in time. Changes in the particle state
can also be caused by the wavefunction reduction, which, as
discussed also in the Introduction, can in principle violate any
feature of dynamic evolution of a quantum system. This can be
expressed quantitatively through the ``Bell'' inequalities
generalizing the classic Bell's inequalities which quantify
violation of the no-action-at-a-distance principle. For
measurements of a mesoscopic qubit (Fig.~\ref{fm2}), the
peculiarities of quantum dynamics of the system originate from the
possibility of quantum-coherent uncertainty in the position of the
charge or flux between the two basis states $|j\rangle$ of the
qubit. In the regime of coherent oscillations of the qubit
($\Delta \neq 0$), this uncertainty gives rise to several
``temporal'' Bell inequalities \cite{q21,q22,q23,q24}. In this
Section, we discuss a sequence of quantum transformation that is
centered around the qubit dynamics with suppressed tunneling,
$\Delta=0$. The transformations lead to the associated Bell-type
inequality, which quantify violation of the fundamental intuition
of many solid-state physicists: charge or magnetic flux can not
tunnel through an infinitely large barrier. The sequence of
transformation includes a measurement done on the qubit and shows
that the wavefunction reduction can indeed violate this
Schr\"{o}dinger-equation-based intuition, and a particle can be
transferred through an ``impenetrable'' barrier in the process of
quantum measurement.

The required manipulations of the qubit state are close to those
in current experiments on coherent oscillations and more complex
dynamics of mesoscopic qubits. Although the discussion in this
Section applies in general to all types of qubits, the physics
content of the wavefunction reduction is more striking for the
semiconductor \cite{q4,q6,q7} or superconductor
\cite{b1,l10,b2,b3,b4} charge qubits, or for the flux qubits
\cite{l11,b5,b6}. In this case, the basic set-up is equivalent to
the one shown in Fig.~\ref{fm2}. The two basis states $|j\rangle
$, $j=1,2$, of the qubit differ by some amount of magnetic flux or
by an individual charge (electron charge $e$ in semiconductor
quantum dot qubit, or Cooper-pair charge $2e$ in a superconductor
qubit) localized on the opposite sides of a tunnel barrier. The
states are coupled by the tunnel amplitude $\Delta >0$. At the
point of resonance, when the bias energy $\epsilon$ between the
two basis states vanishes, the qubit Hamiltonian reduces to
\begin{equation}
H=-\Delta \sigma_x \, , \label{e34}
\end{equation}
and describes quantum coherent oscillations with frequency
$2\Delta$. The tunnel amplitude $\Delta$ and the bias energy
$\epsilon$ are assumed to be controlled externally.

The main element of the sequence of qubit transformations consists
in preparing a superposition (for simplicity, symmetric
superposition, $\sigma_x=1$) of the qubit basis states, then
switching off the tunnel amplitude $\Delta$ and performing
quantum-limited but weak measurement of the qubit position in the
$\sigma_z$ basis. The idea is to prove that the
measurement-induced transfer of the qubit wavefunction between the
two qubit states at $\Delta=0$ gives not only the changes of the
probability (representing our knowledge about the qubit position)
but the actual transfer of charge or flux through the infinitely
large barrier. In order to do this, we include the
measurement-induced wavefunction transfer as a part of the cyclic
transformation, the other part of which is known to transfer
charge or flux. In the ideal situation, when the operations are
precise and there is no intrinsic decoherence, the cycle should
lead to precisely the same initial state $\sigma_x=1$, making it
possible to conclude that the measurement part of it transferred
the qubit state through the barrier with vanishing tunneling
amplitude. In the presence of external perturbations, there will
be a probability $p^{(-)}$ to find the qubit not in the initial
state. One can, however, derive a condition in the form of
inequality on $p^{(-)}$, violation of which shows that this
observation cannot be explained within the assumption of some
initial classical probability distribution over the qubit basis
states. This means that explanation of the observed violation
should necessarily involve the transfer of the qubit state through
suppressed barrier by the process of the wavefunction reduction.

In detail, the starting point of the sequence of transformations
is the ground state of the Hamiltonian (\ref{e34})
\begin{equation}
|\psi_0 \rangle=(|1 \rangle+ |2 \rangle)/\sqrt{2} \, , \label{e35}
\end{equation}
in which the qubit will find itself because of the unavoidable,
but assumed to be weak, relaxation processes, if the Hamiltonian
(\ref{e34}) is kept stationary for some time. Starting from this
state, the tunnel amplitude $\Delta$ is abruptly switched off,
$\Delta=0$. The rate of this process is not important in the case
of the Hamiltonian (\ref{e34}), since the final state will
coincide with (\ref{e35}) regardless of how slowly or quickly
$\Delta$ is switched off. However, in the presence of some
parasitic residual bias $\epsilon$, the rate of variation of
$\Delta$ should be  much larger than $\epsilon$ to preserve the
state (\ref{e35}) at the end of the switching process. Next, as
the first step of actual transformations of the qubit state
(\ref{e35}), the weak quantum-limited measurement of the
$\sigma_z$ operator is performed on this state. The result of this
operation does not depend on the specific model of measurement, as
long as it is quantum-limited. For such a measurement, we know
that specifying the detector output $n$ leaves the qubit in a pure
state which is obtained from (\ref{e35}) by increased degree of
localization in the $\sigma_z$ basis because of the information
about $\sigma_z$ provided by the measurement:
\begin{equation}
|\psi_1 \rangle=\alpha_n |1 \rangle+ \beta_n |2 \rangle \, .
\label{e36}
\end{equation}

\begin{figure}
\hspace{3.5cm} \epsfxsize=5cm \epsfbox{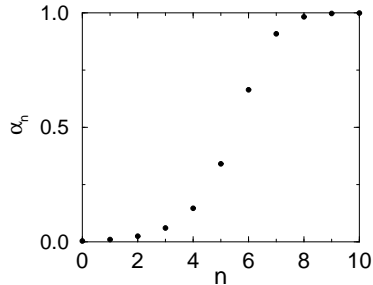}
\caption{Probability amplitude $\alpha_n$ \protect (\ref{e37}) of
finding the qubit in the state $|1\rangle$ as a function of the
observed number $n$ of particles transmitted in $N=10$ attempts
through the ballistic detector with transmission probabilities
$T_1=0.8$ and $T_2=0.4$. The detector measures the state
(\ref{e35}). } \label{fm3}
\end{figure}

As a suitable model of this measurement one can use the
measurement with a ballistic detector discussed above. In this
case, the detector output is the number $n$ of transmitted
particles in $N=ft$ scattering attempts. In this case, according
to Eq.~(\ref{e33})
\[ \alpha_n = \frac{t_1^nr_1^{(N-n)}}{\left[T_1^n R_1^{(N-n)} + T_2^n
R_2^{(N-n)} \right]^{1/2}}\, , \;\;\; \beta_n = \frac{ t_2^n
r_2^{(N-n)} }{ \left[T_1^n R_1^{(N-n)} + T_2^n R_2^{(N-n)}
\right]^{1/2}} \, .\] Using the condition (\ref{e19}) of the
detector ideality one can see that the relative phase $\xi$ of the
coefficients $\alpha_n$ and $\beta_n$ is independent of the
detector output $n$: $\xi=N[\arg(t_1)-\arg(t_2)] $. It can be
viewed then as renormalization of the qubit bias $\epsilon$ due to
the detector-qubit coupling, and can be compensated for by the
bias shift $\delta \epsilon = f [\arg(t_1)-\arg(t_2)]$ during the
period of measurement. The coefficients $\alpha, \beta$ have then
the following form:
\begin{equation}
\alpha_n = \left[ \frac{w_1^{(n)}}{w_1^{(n)} + w_2^{(n)}}
\right]^{1/2} , \;\;\; \beta_n = \left[\frac{ w_2^{(n)}}{
w_1^{(n)} + w_2^{(n)} } \right]^{1/2} , \label{e37}
\end{equation}
where \[ w_j^{(n)}= T_j^n R_j^{(N-n)} \] can be interpreted as the
relative probability for the qubit to be in the state $|j\rangle$
for a given detector outcome $n$. In the situation when the
detector provides no information on $\sigma_z$, e.g. if $T_1=T_2$,
the coefficients (\ref{e37}) are unchanged from their initial
values (\ref{e35}). Otherwise, the probability amplitude is
shifted in the direction of the more probable state: the amplitude
(\ref{e37}) of one qubit state is increased in comparison with
(\ref{e35}) if the observed $n$ is closer to the value $n_j=T_jN$
characteristic for this state than the other state. As an
illustration, Fig.~\ref{fm3} shows the amplitude $\alpha_n$ for
$N=10$, $T_1=0.8$, and $T_2=0.4$, as a function of $n$. One can
see that $\alpha_n$ maintains its original value $1/\sqrt{2}$ from
Eq.~(\ref{e35}) for $n$'s roughly in the middle between the two
characteristic values $n_1=8$ and $n_2=4$. For $n$ smaller than or
close to $n_2$, $\alpha_n$ decreases from its original value, for
$n$ close to or larger than $n_1$, $\alpha_n$ approaches 1. Such a
shift due to the wavefunction reduction is the central part of the
transformation cycle.

\begin{figure}
\hspace{2.4cm} \epsfxsize=8cm \epsfbox{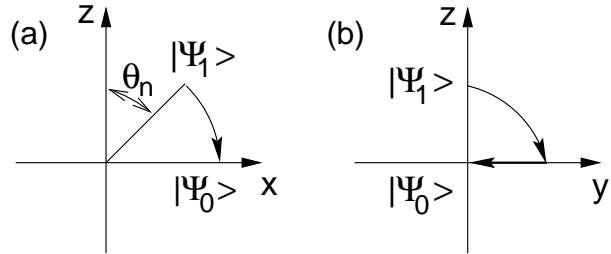} \caption{Diagram
of the two possible transformations of the qubit state in the spin
representation  returning the state $|\psi_1 \rangle$ to $|\psi_0
\rangle$ after the measurement-induced state reduction $|\psi_0
\rangle \rightarrow |\psi_1 \rangle$. (a) Direct one-step $y$-axis
rotation \protect (\ref{e38}). (b) Projection on the $z-y$ plane
of the two-step transformation (\ref{e40}) with the same end
result. } \label{fm4}
\end{figure}

The remaining steps of the cycle aim at returning the qubit to its
initial state (\ref{e35}). To do this, one needs to transfer back
the charge or flux that was transferred in the wavefunction
reduction process leading to the state (\ref{e37}). This is
achieved by creating for some time the non-vanishing tunneling
amplitude, i.e. realizing a fraction of a period of the regular
coherent oscillations in which the charge or flux goes
back-and-forth between the two qubit basis states. In the most
direct way, this can be done if the qubit structure makes it
possible to create non-vanishing phase of the tunnel amplitude
$\Delta'(t)$ (e.g., in the superconducting qubits, where the
tunnel amplitude can be controlled through quantum interference,
producing any complex value of this amplitude). In this case, the
state (\ref{e37}) can be returned back directly into the initial
form (\ref{e35}) if $\arg{\Delta'}=\pi/2$. In the diagram
(Fig.~\ref{fm4}a) in which the qubit states are represented in the
language of spin-1/2, i.e. \[ |\psi_1 \rangle=\cos (\theta_n/2) |1
\rangle+ \sin (\theta_n/2) |2 \rangle \, , \] such a tunneling
amplitude corresponds to rotation around the $y$ axis. The diagram
in Fig.~\ref{fm4}a shows then directly that the rotation around
$y$ axis turning $|\psi_1 \rangle$ into $|\psi_0 \rangle$ should
have the magnitude:
\begin{equation}
\int |\Delta'(t)|dt/\hbar= (\pi/2- \theta_n)/2 \, , \label{e38}
\end{equation}
where
\[\theta_n = 2 \tan^{-1}(\beta_n/\alpha_n) =
2 \tan^{-1}\left[(T_2/T_1)^{n/2}(R_2/R_1)^{(N-n)/2}\right] .\]

If the qubit structure allows only for the real tunnel amplitude
$\Delta$ (the situation that can be expected, e.g., in
semiconductor quantum dot qubits), the $y$-axis rotation $R_y=
\exp \{-i \sigma_y \int |\Delta'(t)|dt/\hbar \}$ (\ref{e38}) can
be simulated in three steps in which the rotation $R_x= \exp \{-i
\sigma_x \int \Delta(t)dt/\hbar\}$ around the $x$ axis of the same
magnitude (\ref{e38}) is preceded and followed by the rotations
around the $z$-axis:
\begin{equation}
R_y=U^{-1} R_x U\, , \;\;\; U=\exp \{i \sigma_z \pi/4 \}\, .
\label{e39}
\end{equation}
The $z$-axis rotations can be created by the pulses of the qubit
bias: $\int \epsilon (t)dt/\hbar = \pm \pi/4$. The three-step
sequence (\ref{e39}) can be simplified into two steps
(Fig.~\ref{fm4}b) by changing the order of rotations: first, the
$x$-axis rotation by $\pi/4$ (opening tunneling $\Delta(t)$ for
appropriate interval of time) followed by one $z$-axis rotation:
\begin{equation}
\int \Delta(t)dt/\hbar= \pi/4 \, , \;\;\;\;\; \int \epsilon
(t)dt/\hbar = (\pi/2- \theta_n)/2 \, . \label{e40}
\end{equation}

All these versions of the transformation cycle bring the qubit to
its initial state (\ref{e35}). In all cases, completion of the
cycle that started with a shift of the wavefunction amplitudes due
to the state reduction involves a part of a period of coherent
oscillations which reverses this shift. Coherent qubit
oscillations are known to actually transfer the charge of flux
between the two qubit states. Since the cycle as a whole is
closed, this fact shows that the changes in the qubit state caused
by the wavefunction reduction can not be interpreted only as the
changes in our knowledge of probabilities of the state of the
qubit, but involve the actual transfer of the charge or flux in
the absence of the tunneling amplitude.

To see this more quantitatively, one can derive the Bell-type
inequality, violation of which should show that understanding of
the state reduction solely in terms of probability changes can not
be correct. The inequality is obtained by assuming that the
process of switching off the tunneling amplitude $\Delta$ in the
beginning of the transformation cycle does not leave the state
(\ref{e35}) unchanged but instead localizes the qubit in one of
the basis states. This means that the process produces an
incoherent mixture of the qubit states with some, in general
unspecified, probability $p$ to be in the state $|1\rangle$. This
process would provide then an alternative, classical description
of the evolution during the measurement process. In this
description, the qubit state is ``objectively'' well defined, but
is unknown to us, and the measurement gradually provides
information about this unknown state. The measurement would only
change the probabilities we ascribe to the two qubit states, but
not the state itself, and in particular would not transfer the
charge of flux. One should then see how well this classical
description can mimic the quantum result of the transformation
cycle described above. A convenient way of making this comparison
is provided by the probability of ending the cycle in the wrong
state. The unperturbed quantum evolution should end up in the
initial state $\sigma_x=1$, whereas the same transformation cycle
performed on the classical initial state will always have a finite
probability $p^{(-)}$ of ending in the state $\sigma_x=-1$.

This probability is found by applying the transformations not to
the state (\ref{e35}) but to the incoherent state with the density
matrix
\[ \rho_0 =p|1\rangle \langle 1| + (1-p)|2\rangle
\langle 2|\, . \] In this case, the measurement changes only the
probability $p$ in this expression. Similarly to Eq.~(\ref{e37}),
if the detector gives the output $n$, the density matrix of the
system is
\begin{equation}
\rho_1= \rho_0 \big|_{p\rightarrow \bar{p}}\, , \;\;\;\; \bar{p}=
\frac{pw_1^{(n)}}{pw_1^{(n)} + (1-p)w_2^{(n)} }\, . \label{e41}
\end{equation}
All versions (\ref{e38}) -- (\ref{e40}) of the transformation
cycle produce the same probability $p^{(-)}$ of being in the state
$\sigma_x=-1$, when applied to the density matrix $\rho_1$
(\ref{e41}). For instance, one can see directly that in the
density matrix $R_y^{-1}\rho_1R_y$ obtained from $\rho_1$ by
rotation (\ref{e38}), the probability $p^{(-)}$ is:
\[ p^{(-)}=\frac{w_1w_2}{(w_1 + w_2)(pw_1 + (1-p)w_2) } \, .\]

Minimizing this expression with respect to $p$, we see that the
minimum probability of finding the state $\sigma_x=-1$ in the
classical case is
\begin{equation}
p^{(-)}= \frac{\mbox{min} \{ w_1\, ,w_2\} }{w_1 + w_2 }\, .
\label{e42}
\end{equation}
Instead of looking for the minimum with respect to $p$, one can
adopt a natural additional assumption that when the tunneling
amplitude is switched off, the qubit localization process can only
be symmetric, since there is no reason to prefer one qubit state
to another. In this case, $p=1/2$, and we obtain somewhat
different expression for the probability $p^{(-)}$ with
qualitatively similar properties:
\begin{equation}
p^{(-)}= \frac{2 w_1w_2 }{(w_1 + w_2)^2 }\, . \label{e43}
\end{equation}
This expression would also be obtained if the qubit wavefunction
would be reduced to the density matrix $\rho_1$ (\ref{e41}) during
the measurement, not in the process of suppression of the
tunneling amplitude.

Equations (\ref{e42}) and (\ref{e43}) show that in order to
distinguish the quantum coherent evolution (for which $p^{(-)}=0$)
and incoherent evolution with non-vanishing probability $p^{(-)}$,
it is important to employ weak measurement. If the measurement is
projective, i.e. if one of the probabilities $w_j$ is zero so that
the measurement completely reduces the qubit state to one of the
basis states, then $p^{(-)}=0$ and it is impossible to distinguish
the two types of evolution. This conclusion should be independent
of the specific form of the employed transformation cycle, since
projective measurement is always expected to fully separate
different components of the initial state of the measured system
and completely suppress quantum coherence between them.

The discussion above means that observation of the probability of
the state $\sigma_x=-1$ smaller than $p^{(-)}$,
\begin{equation}
p(\sigma_x=-1) < p^{(-)} \label{e44}
\end{equation}
at the end of the transformation cycle proves that all
transformations in this cycle, including the wavefunction
reduction, are quantum coherent. Combined with the non-vanishing
transfer of charge or flux during the ``oscillation'' step
[(\ref{e38}) -- (\ref{e40})] of the cycle, this fact implies that
the wavefunction reduction induces similar transfer across the
tunnel barrier separating the qubit basis states even if the
corresponding tunnel amplitude is zero.

It is important to note that this counter-intuitive feature of the
wavefunction reduction does not contradict the fact that all
dynamic properties of the measurement, including the back-action
dephasing and information rates can be calculated from the dynamic
detector model without any reference to the wavefunction
reduction. While the dynamic properties are average
characteristics of the detector, the wavefunction reduction
appears if one considers separately individual outcomes of
measurement. In the transformations discussed above, this
separation is achieved by introducing the feedback, operations on
the measured system which depend on the specific measurement
outcome. In ballistic detectors, the measurement outcomes are
distinguished by the number $n$ of transmitted particles, and to
see the wavefunction reduction one needs to distinguish individual
particles, and is done, e.g., in the electron counting experiments
\cite{q25}. For the detectors for which distinguishing individual
transmitted particles can be problematic (e.g., the QPC detector),
the allowed uncertainty in $n$ should be smaller than the width
$\delta n$ of the transition region in the $n$-dependence of the
wavefunction amplitudes of the measured qubit -- see
Fig.~\ref{fm3}. Since the width of this region can be estimated
roughly as $\delta n \simeq 1/\ln(T_1R_2/T_2R_1)$, the uncertainty
in $n$ can be compensated for by making the difference between
transmission probabilities $T_j$ smaller, thus increasing $\delta
n$. The limit to this increase is set by the decoherence processes
in the measured system which make it impossible to increase the
measurement time of the detector beyond the coherence time of the
system without losing the non-trivial character of the measurement
dynamics.

\section{Tunneling detectors}

So far, the discussion was based on the ballistic model of the
mesoscopic detector, in which the measured system controls
ballistic motion of some particles between the two reservoirs
(Fig.~\ref{fm1}). If one assumes that the particle transmission
probabilities are small, $T_j \ll 1$, and the transfer processes
between the reservoirs can be described in the tunneling
approximation, specific nature of the detector transport becomes
irrelevant. In this case, the range of applicability of the
detector model can be extended significantly to include the
detectors in which it is not possible to identify regions of
ballistic transport, but which are still based on the very similar
dynamic principle: control by the measured system of transport
between the two reservoirs. Examples of such ``tunneling''
detectors include the superconducting SET electrometer
\cite{b7,b8,b9}, normal SET electrometer in the co-tunneling
regime \cite{b14}, or dc SQUID magnetometer (see, e.g.,
\cite{b10,b11,b12,b13}) used for measurements of superconducting
qubits. The aim of this Section is to show briefly that the
measurement properties of this type of mesoscopic detectors
coincide in essence with those of the ballistic detectors.

Since the measured system controls the tunneling amplitude
$\hat{t}$ of particles in the detector, this amplitude should be
treated as a non-trivial operator acting on the measured system.
The detector tunneling can be described then with the standard
tunnel Hamiltonian, the transfer terms in which are split into a
product of operators of the measured system and the detector. In
this case, the tunnel Hamiltonian describes the detector-system
coupling and can be written as
\begin{equation}
H_T=\hat{t} \xi+\hat{t}^{\dagger}  \xi^{\dagger} \, , \label{e1}
\end{equation}
where $\xi, \xi^{\dagger}$ are the operators that describe the
detector part of the tunneling dynamics, e.g., creation of
excitations in the detector reservoirs when a particle tunnels,
respectively, forward and backward between them. Inclusion of the
operators $\xi, \xi^{\dagger}$ means, therefore, that the
Hamiltonian (\ref{e1}) makes it possible to describe the detectors
in which the tunneling processes are strongly inelastic. This
fact, however, does not prevent correct account of elastic
transport in the case of ballistic detectors. Qualitative reason
for this can be seen easily using as an example the QPC detector.
While scattering of individual electrons in the QPC is elastic, in
the tunnel limit, electron transfer between the two electrodes of
the QPC can also be viewed as creation of electron-hole excitation
in the electrodes, with an electron removed from a state below the
Fermi level in one electrode, and transferred to a state above the
Fermi level in the other. Accordingly, as we will see later,
Hamiltonian (\ref{e1}) leads to the evolution equations that
coincide in the tunnel limit with those obtained above in the
ballistic case.

Under the assumption that the detector tunneling is weak, the
precise form of the internal detector Hamiltonian is not important
and dynamics of measurement is defined by the correlators of the
operators $\xi, \xi^{\dagger}$:
\begin{equation}
\gamma_+=\int_{0}^{\infty} dt\langle \xi(t) \xi^{\dagger}\rangle
\, , \;\;\; \gamma_-=\int_{0}^{\infty}dt\langle \xi^{\dagger}(t)
\xi\rangle \, . \label{e3}
\end{equation}
Here the angled brackets denote averaging over the detector
reservoirs which are taken to be in a stationary state with some
fixed number of particles in them and the density matrix $\rho_D$:
$\langle ... \rangle = \mbox{Tr}_D \{ ... \rho_D \}$. The
correlators (\ref{e3}) set the scale $\Gamma_{\pm}\equiv
2\mbox{Re} \gamma_{\pm}$ of the forward and backward tunneling
rates in the detector.

A reasonable tunneling detector should satisfy some additional
assumptions related to the fact that its output should be
classical in order to provide a complete measurement dynamics.
Similarly to the ballistic detector, the output information in the
tunneling case is contained in the number $n$ of the particles
transmitted between the detector reservoirs. For this number to
behave classically, the correlators $\langle \xi(t) \xi \rangle$,
$\langle \xi^{\dagger} (t) \xi^{\dagger} \rangle$ that do not
conserve the number of tunneling particles should vanish. Another
consequence of the assumption of classical detector output is that
the energy bias $\Delta E$ for tunneling through the detector
should be much larger than the typical energies of the measured
system. In this case, one can neglect quantum fluctuations of the
detector current in the relevant frequency range that corresponds
to frequencies of evolution of the measured system. If one assumes
in addition that all other characteristic frequencies of the
detector tunneling are also much larger than those of the measured
system, the functions $\xi(t)$, $\xi^{\dagger}(t)$ in
Eq.~(\ref{e3}) are effectively $\delta$-correlated on the time
scale of the system.

Vanishing correlation time in the correlators (\ref{e3}) makes it
possible to write down simple evolution equations for the density
matrix $\rho$ of the measured system. To describe the system
dynamics conditioned on particular outcome of measurement, we also
keep in the evolution equation the number $n$ of particles
transferred through the detector. Since the correlators that do
not conserve $n$ vanish, only the terms diagonal in $n$ are
important. In the interaction representation with respect to the
tunnel Hamiltonian (\ref{e1}), the density matrix $\rho(t)$ is
given by the standard expression:
\begin{equation}
\rho(t) = \langle S \rho \rho_D S^{\dagger} \rangle , \;\; S = T
\exp \{-i\int^tdt' H_T (t') \} . \label{e5}
\end{equation}
For $\delta$-correlated operators in Eq.~(\ref{e3}), one can see
that the full perturbation expansion of Eq.~(\ref{e5}) in $H_T$ is
equivalent to the evolution equation  for $\rho(t)$ that follows
from the lowest-order perturbation theory. Keeping track of the
number $n$ of particles transferred through the detector, we get:
\begin{equation}
\;\;\;\; \dot{\rho}^{(n)}= \Gamma_+\,
\hat{t}^{\dagger}\rho^{(n-1)} \hat{t}+\Gamma_- \, \hat{t}
\rho^{(n+1)} \hat{t}^{\dagger} -(\gamma_+\hat{t}\hat{t}^{\dagger}
+ \gamma_-\hat{t}^{\dagger}\hat{t})\rho^{(n)} -
\rho^{(n)}(\gamma^*_+\hat{t} \hat{t}^{\dagger}+ \gamma^*_-
\hat{t}^{\dagger}\hat{t}) \, . \label{e4}
\end{equation}

Since the tunneling amplitude $\hat{t}$ is a function of some
observable of the measured system, there is a system of
eigenstates $|j\rangle$ common to the operators $\hat{t}$ and
$\hat{t}^{\dagger}$:
\[ \hat{t}|j\rangle = t_j |j\rangle\, , \;\;\;\;
\hat{t}^{\dagger}|j\rangle = t_j^* |j\rangle\, , \] where $t_j$ is
the detector tunneling amplitude when the measured system is in
the state $|j\rangle$. It is convenient to write the evolution
equation (\ref{e4}) in the basis of states $|j\rangle$:
\begin{eqnarray}
\dot{\rho}^{(n)}_{ij}= -(1/2)(\Gamma_+ +\Gamma_-)(|t_i|^2+
|t_j|^2) \rho^{(n)}_{ij} + \Gamma_-\, t_it_j^*\rho^{(n+1)}_{ij}
\label{e6} \\ + \Gamma_+ \, t_i^*t_j \rho^{(n-1)}_{ij}  - i[\delta
H, \rho^{(n)}]_{ij} \, , \nonumber
\end{eqnarray} where
\begin{equation}
\delta H = \mbox{Im}(\gamma_+ +\gamma_-) |t_j|^2\, |j\rangle
\langle j| \label{ren}
\end{equation}
is the renormalization of the Hamiltonian of the measured system
due to its coupling to the detector.

If one omits the term $\delta H$ which can be combined with the
internal Hamiltonian of the measured system, Eq.~(\ref{e6}) can be
solved in $n$ by noticing that it coincides in essence with a
recurrence relations for the modified Bessel functions
$I_n$~\cite{q26}. In order to interpret $n$ as the number of
particles transferred through the detector during the time
interval $t$, we solve this equation with the initial condition
$\rho^{(n)}_{ij}(t=0)= \rho_{ij}(0) \delta_{n,0}$. The
corresponding solution is:
\begin{equation}
\;\;\; \frac{\rho^{(n)}_{ij}(t)}{\rho_{ij}(0)}= \left(
\frac{\Gamma_+}{\Gamma_-}\right)^{n/2} I_n
(2t|t_it_j|\sqrt{\Gamma_+\Gamma_-}) \exp \{ -\frac{\Gamma_+
+\Gamma_-}{2} (|t_i|^2+|t_j|^2) t -in \varphi_{ij} \} \, ,
\label{e7}
\end{equation}
where $\varphi_{ij} \equiv \arg(t_it_j^*)$. As discussed above,
the qubit density matrix conditioned on the particular measurement
outcome $n$ follows from Eq.\ (\ref{e7}) if one selects in this
equation the terms with given $n$ and normalizes the resulting
reduced density matrix back to 1. If one of the tunneling rates
$\Gamma_+$ or $\Gamma_-$ vanishes, Eq.\ (\ref{e7}) reduces to the
usual Poisson distribution characteristic for tunneling in one
direction. Conditional description of measurement in this case was
developed in \cite{q11}. When both rates are non-vanishing,
specifying the total number $n$ of the transferred particles does
not specify uniquely evolution of the detector, since the same $n$
results from the balance between different numbers of particles
transferred forward and backward. This means that some information
is lost in this regime and the detector is not quantum-limited
(see the discussion below). In contrast to the situation with the
quantum-limited detectors considered in the previous Sections,
conditional dynamics that follows from Eq.~(\ref{e7}) in this case
\cite{q27} does not preserve the purity of the quantum state of
the measured system.

Evolution of the density matrix $\rho$ averaged over the
measurement outcomes $n$ can be obtained by either disregarding
simply the index $n$ in Eq.~(\ref{e6}), or directly taking the sum
over $n$ in (\ref{e7}) with the help of a summation formula
\cite{q26} for the Bessel functions. In both cases, equation for
the measurement-induced evolution of $\rho$ is:
\begin{equation}
\dot{\rho}_{ij}= -\Gamma_{ij}\rho_{ij} -i (\Gamma_+
-\Gamma_-)|t_it_j| \sin \varphi_{ij}\rho_{ij}\, , \label{e8}
\end{equation}
where
\begin{equation}
\Gamma_{ij} \equiv (1/2)(\Gamma_+ +\Gamma_-)|t_i-t_j|^2 \label{e9}
\end{equation}
is the back-action dephasing rate by the tunneling detector. The
last term in Eq.~(\ref{e8}) can be viewed as another contribution
to renormalization of the energy difference between the states
$|i\rangle$ and $|j\rangle$, although in general it can not be
reduced to the energy shifts of individual states [in contrast to
the renormalization term (\ref{ren})].

The back-action dephasing rate (\ref{e9}) coincides with that of
the ballistic detectors in the tunneling limit given by
Eq.~(\ref{e151}), if we take into account that our discussion of
ballistic detectors assumed for simplicity that particles are
incident on the scattering region from only one electrode. It can
be seen directly that the difference in the phases [see
Eq.~(\ref{e152})] of the tunneling amplitudes in the two
expressions is not essential. The reason for this difference is
that the tunnel Hamiltonian (\ref{e1}) describes explicitly only
the effects associated with the actual transfer of particles
across the detector. It is assumed that the other effects of the
detector-system coupling, e.g., renormalization of the system
energy due to reflection processes in the detector, are already
accounted for. In terms of the scattering amplitudes, this implies
that the reflection amplitudes $r$ and $\bar{r}$ for particles
incident on the tunnel barrier from the two electrodes of the
detector, should satisfy the conditions $\arg(r_ir_j^*)=0$ and
$\arg(\bar{r}_i\bar{r}_j^*)=0$. In this case, Eq.~(\ref{e151})
coincides with (\ref{e9}) even with account of extra phases
(\ref{e152}). This means that the model of the tunnel detector
considered in this Section is equivalent to that of the ballistic
detector in the appropriate small-transparency limit.

\begin{figure}
\hspace{3.5cm} \epsfxsize=5.5cm \epsfbox{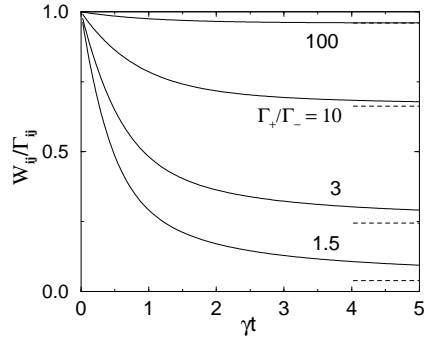} \caption{The
information acquisition rate $W_{ij}$ of the tunneling detector
normalized to the dephasing rate $\Gamma_{ij}$ \protect (\ref{e9})
(for $\varphi_{ij}=0$) as a function of time $t$ for several
ratios of the forward and backward tunneling rates. The time $t$
is normalized to the typical forward tunneling rate
$\gamma=\Gamma_+\, (T_i+T_j)/2$. The curves are plotted for the
detector transparencies $T_i=0.2$, $T_j=0.4$. The dashed lines
show the corresponding asymptotic values (\ref{e51}). For
$\Gamma_+/\Gamma_-=100$, the dashed line overlaps with the main
curve. } \label{fm5}
\end{figure}

To calculate the information rate $W_{ij}$ of the tunneling
detector in the situation when the rates of both forward and
backward tunneling are non-vanishing (in contrast to scattering
from one direction discussed for the ballistic detectors), one
needs to use in Eq.~(\ref{e17}) the diagonal part of
Eq.~(\ref{e7}) which gives the probabilities $p_j(n) =
\rho^{(n)}_{jj}$ for $n$ particles to tunnel when the system is in
the state $|j\rangle$. An example of the rate $W_{ij}$ defined in
this way is shown in Fig.~\ref{fm5}. This figure shows that in
general $W_{ij}$ is time-dependent and approaches constant value
only after a transition period. If the tunneling probabilities
$T_j=|t_j|^2$ do not differ very strongly, this transition period
is shorter than the back-action dephasing time $\Gamma_{ij}^{-1}$.

The constant asymptotic values of the information rate can be
obtained from the asymptotic behavior of the Bessel functions
$I_n(z)$. Using the standard integral representation for $I_n(z)$
in Eqs.~(\ref{e17}) and (\ref{e7}), and making use of the fact
that the constant rates $W_{ij}$ are determined by the exponential
behavior of the integrals at large time $t$, we find directly
\begin{equation}
W_{ij}=(\Gamma_+ +\Gamma_-)(T_i+T_j)/2 -
\left[(T_i^2+T_j^2)\Gamma_+\Gamma_- + T_iT_j (\Gamma_+^2
+\Gamma_-^2)\right]^{1/2} \, .  \label{e51}
\end{equation}
If the particles tunnel only in one direction, e.g. $\Gamma_- =0$,
Eq.~(\ref{e51}) reduces to the previously known result
\cite{gur,q19} $W_{ij}=\Gamma_+ (\sqrt{T_i}-\sqrt{T_j})^2/2$, in
which the information rate and the back-action dephasing rates
coincide, when the phases of the tunneling amplitudes satisfy the
appropriate ideality condition $\varphi_{ij}=0$. In the other
limit of small difference $2\Delta T$ between the transmission
probabilities $T_{i,j}= T\pm \Delta T $, Eq.~(\ref{e51}) reduces
to
\begin{equation}
W_{ij}=\frac{(\Delta T)^2(\Gamma_+ - \Gamma_-)^2}{2 T (\Gamma_+
+\Gamma_-)} \, .  \label{e52}
\end{equation}
This equation agrees with the general theory of linear
measurements (see, e.g., \cite{q10}), in which it can be
interpreted as the rate with which one can distinguish the
difference $\Delta T(\Gamma_+ - \Gamma_-)$ between the two
detector currents in the presence of the current noise with
spectral density $T (\Gamma_+ +\Gamma_-)$.

In the most typical situation, the two rates $\Gamma_{\pm}$ are
both non-vanishing because of the finite temperature $\Theta$ of
the detector electrodes. The electrodes can be in equilibrium even
when a non-vanishing current is driven between them by finite
energy difference $\Delta E$ created for the tunneling particles.
In this case, the tunneling rates $\Gamma_{\pm}$ are related by
the detailed balance relation and can be written as $\Gamma_{\pm}=
\Gamma_0 \exp (\pm \Delta E/2\Theta)$, where $\Gamma_0$ is the
typical tunneling rate which can also depend on temperature and
energy bias. The information rate (\ref{e51}) then is
\begin{equation}
W_{ij}/\Gamma_0 =(T_i+T_j)\cosh (\Delta E/2\Theta )- \left[
T_i^2+T_j^2+ 2T_iT_j \cosh (\Delta E/\Theta )\right]^{1/2} \, .
\label{e53}
\end{equation}
Comparison of Eqs.~(\ref{e53}) or (\ref{e51}) with Eq.~(\ref{e9})
for the back-action dephasing rate (see also Fig.~\ref{fm5}) shows
that the detector with temperature $\Theta \sim \Delta E$ which
creates non-vanishing rates $\Gamma_{\pm}$, is not
quantum-limited, $W_{ij} < \Gamma_{ij}$, even if the phases of the
tunneling amplitudes satisfy the ideality condition
$\varphi_{ij}=0$. Equation (\ref{e53}) can be used to establish
quantitative condition on the detector temperature necessary for
the desired degree of the detector ideality.

\section{Conclusion}

Two general models of realistic mesoscopic solid-state detectors
have been described in this paper. The detectors are based on
ability of the measured system to control transport current
between two particle reservoirs. The models enable detailed
analysis of the dynamics of the measurement process. Wavefunction
reduction is introduced in this dynamics through the assumption of
suppressed quantum coherence between the particle states in
different reservoirs. This procedure is very natural and can be
justified qualitatively within the general approach to decoherence
in quantum systems. The main element of the justification is the
increased level of difficulty of maintaining quantum coherence
between the states of progressively more complex systems. This
fact makes the boundary between the quantum and classical domains
not very sharp and dependent on details of specific measurement
set-up. This leads to an interesting question whether it is
possible to formulate more general and self-consistent conditions
defining the boundary between the quantum and classical behaviors
of dynamic systems. Mesoscopic solid-sate structures provide a
convenient setting for further studies of this question.

\vspace{6ex}

\acknowledgments

Part of the discussion in this paper is based on the work done in
collaboration with A. di Lorenzo, A.N. Korotkov, K. Rabenstein, R.
Ruskov, V.K. Semenov, E.V. Sukhorukov, and W. Mao. The author
would like to thank them, and also G.~Benenti, R.~Fazio, J.W. Lee,
F.~Plastina, and D.L.~Shepelyansky, for collaboration and
discusssions of quantum measurements. This work was supported by
the NSF under grant \# DMR-0325551.

\end{document}